# Flexo-Sensitive Polarization Vortices in Thin Ferroelectric Films

Anna N. Morozovska[1], Eugene A. Eliseev[2], Sergei V. Kalinin[3*], and Riccardo Hertel[4†],

[1] Institute of Physics, National Academy of Sciences of Ukraine, 46, pr. Nauky, 03028 Kyiv, Ukraine

[2] Institute for Problems of Materials Science, National Academy of Sciences of Ukraine, Krjijanovskogo 3, 03142 Kyiv, Ukraine

[3] The Center for Nanophase Materials Sciences, Oak Ridge National Laboratory, Oak Ridge, TN 37831

[4] Université de Strasbourg, CNRS, Institut de Physique et Chimie des Matériaux de Strasbourg, UMR 7504, 67000 Strasbourg, France

## Abstract

We consider theoretically the influence of the flexoelectric coupling on the spatial distribution and temperature behavior of spontaneous polarization for several types of stable domain structure in thin ferroelectric films, such as stripe domains and vortices. Finite element modelling (FEM) for $BaTiO_3$ films and analytical calculations within the Landau-Ginzburg-Devonshire approach reveals that an out-of-plane polarization component can be very sensitive to the flexoelectric coupling for periodic quasi-2D stripe domains and 3D vortex-antivortex structures. However, the influence is rather different for these structures. The flexoelectric coupling increases significantly the amplitude of a small out-of-plane polarization component in the stripe domains, and the "up" or "down" direction of the component is defined by the sign of the flexoelectric coefficients. Concerning the vortex-antivortex pairs, their antivortices with in-plane anti-circulation have smooth wide dipolar cores through the entire film, whose shape and other features are almost insensitive to the coupling. The vortices with in-plane vorticity have spike-like cores with an out-of-plane quadrupolar moment induced by the flexoelectric coupling. The cores are located near the film-dead layer interfaces. FEM results corroborated by analytical calculations prove that a change of the flexoelectric coefficient sign leads to a reorientation of the core axial polarization, making the flexo-sensitive 3D vortices similar to the recently introduced "flexons" in cylindrical nanoparticles. The relatively wide temperature range (from 200 to 400 K) of the flexo-sensitive vortices' existence gives us the hope that they can be observed experimentally in thin ferroelectric films by scanning probe and nonlinear optical microscopy methods.

---

[*] Corresponding author: sergei2@ornl.gov

[†] Corresponding author: riccardo.hertel@ipcms.unistra.fr

# I. Introduction

Since its appearance and until now, nanoscale ferroics (ferromagnets, ferroelectrics, ferroelastics) have been the main object of fundamental research on the physical nature of long-range order of polar, magnetic, and structural properties [1, 2, 3]. The leading role is played by the emergence of a domain structure of long-range order parameters, such as electric polarization, magnetization, and (anti)ferrodistortion, and its interaction with the surface of a nanoferroic [4, 5].

In the case of ferroelectrics, which we discuss here, the long-range order of the domains is governed by both electrostatic fields and the distribution of strain fields and their gradients. A strong and spatially extended gradient of elastic fields generated by the surface, domain walls, and/or other inhomogeneities creates a sufficiently large flexoelectric polarization, which is proportional to this gradient (this phenomenon is a direct flexoelectric effect) [6]. On the other hand, the electrical polarization gradient gives rise to an inhomogeneous strain (inverse flexoelectric effect) [7]. The thermodynamic description of the flexoelectric effect is given by Lifshitz invariants [8, 9].

The flexoelectric effect exists in all ferroics [6-9], in contrast to the flexomagnetic [10] and flexomagnetoelectric [11, 12] effects, the existence of which is critically sensitive to the presence of time inversion and its connection with other operations of point symmetry of a particular material [13, 14]. The influence of the flexoelectric effect on the macroscopic properties of ferroics (as well as any other materials) is relatively small [6-9]. However, in nanosized and nanostructured ferroics, the influence of the flexoelectric effect on their polar, magnetic, electronic properties and phase transitions can be significant and may lead to fundamental changes in their properties [15, 16, 17]. This size-dependence of the effect arises from the gradients of physical quantities playing a leading role in such nanosystems [18, 19].

Recently, it was predicted theoretically that a decrease in the correlation-gradient polarization energy leads to a significant increase in the polarization gradient, which in turn leads to spontaneous bending of otherwise uncharged domain walls in thin multiferroic films [20] and ferroelectric nanoparticles [21]. Such domain walls can form meandering [20] and/or labyrinthine [21] structures. Later, similar structures were discovered experimentally in thin $BiFeO_3$ and $Pb(Zr_{0.4}Ti_{0.6})O_3$ films by HR-STEM [22] and PFM [23] methods, respectively, and corroborated by ab initio calculations [23]. However, the influence of flexo-effects on the morphology of domain structures in thin films has not yet been studied systematically.

Most published experimental studies of flexoelectric phenomena in proper [24, 25, 26] and incipient [27, 28] ferroelectrics, atomistic quantum-mechanical [29, 30] and first-principles calculations [31, 32, 33, 34] are overwhelmingly aimed at determining the magnitude of the flexoelectric coefficients [28-34] and the structure of the flexoelectric tensor [35, 36]. The majority of the phenomenological papers on this topic are devoted to the influence of the flexo-effects on the macroscopic properties and phase transitions of the ferroic film as a whole (see e.g. Refs.[37, 38, 39] and refs therein), and only a few of them address the actual influence of the flexoelectric effect on the structure of domain walls [40, 41, 42, 43, 44]. The main result of the papers [40 - 44] is the prediction that the flexoelectric effect induces the appearance of a small but sufficiently strong polarization component perpendicular to the plane of the nominally uncharged domain wall of the ferroic, resulting in an effective "charging" of the wall. PFM and cAFM experiments registering the conductivity of nominally uncharged domain walls in ferroelectrics and multiferroics confirm the theoretical predictions [45, 46, 47]. Non-Ising and chiral ferroelectric domain walls have been revealed by nonlinear optical microscopy [48].

Note that only a few works analyze the influence of the flexoelectric coupling on the curled vortex-like domain structures in ferroelectric thin films [49] and nanoparticles [50, 51]. Namely, using machine learning and phase-field modeling, Li et at. [49] analyze the role of flexoelectricity on polar 2D vortices in a $PbTiO_3/SrTiO_3$ superlattice. The axis of these vortices is parallel to the film surface, and the flexoelectric coupling influences their details in a quantitative way, but not sharply or critically. The influence of flexoelectricity on polarization vortices in spherical [50] and cylindrical [51] core-shell ferroelectric nanoparticles can be significant because it can induce a small axial polarization of the vortex core. Analytical calculations and simulations [51] have proven that a change of the flexoelectric coefficient sign leads to a reorientation of the axial polarization of the vortex and that an anisotropy of the flexoelectric coupling critically influences the vortex core formation and its related domain morphology. Here we consider the influence of the flexoelectric coupling on the spatial distribution and temperature behavior of the spontaneous polarization for several types of stable domain structures in thin $BaTiO_3$ films, such as periodic quasi-2D stripe domains and arrays of 3D vortex-antivortex pairs, whose axes are perpendicular to the film surface.

## II. Problem Statement

Using a finite element modelling (FEM) and Landau-Ginzburg-Devonshire (LGD) phenomenological approach combined with electrostatic equations and elasticity theory, we model the polarization, internal electric field, elastic stresses, and strains in a thin $BaTiO_3$ film. We consider a $BaTiO_3$ film [001] sandwiched between two ultra-thin paraelectric dead layers with a high relative dielectric permittivity. A top electrode and conducting substrate are in contact with the layers, and the voltage is applied between the conductors (see **Fig. 1a**). The dead layers are required for the thermodynamic stability of the domain structure. Only a single-domain distribution is stable in the case of perfect electric contact between the ferroelectric film surfaces and ideal conducting electrodes. The film thickness varies from 4 to 20 nm, and the thickness of each dead layer varies from 0 to 0.8nm. The relative dielectric permittivity is 300.

For most cases, we impose a slight tensile strain $0 < u_m < 0.25\%$ at the interface BaTiO$_3$ film – substrate. The strain leads to the disappearance of the metastable orthorhombic phase and shifts the temperatures of structural and polar transitions [52]. Also, the strain supports a rhombohedral ferroelectric phase in a single-domain film at temperatures below 360 K. At the same conditions, the formation of in-plane domains is favorable in the film due to depolarization effects. The transition from the rhombohedral to the tetragonal phase occurs at a temperature above 360 K, while the transition from the tetragonal ferroelectric to the paraelectric cubic phase occurs above 420 K (see **Fig. 1b**).

An LGD free energy functional $G$ of the BaTiO$_3$ film includes a Landau energy – an expansion on 2-4-6 powers of the polarization components $P_i$, $G_{Landau}$; a polarization gradient energy, $G_{grad}$; the electrostatic energy, $G_{el}$; an elastic, electrostriction contribution $G_{es}$, a flexoelectric contribution, $G_{flexo}$; and a surface energy term, $G_S$ [37, 39]. The free energy functional, the Euler-Lagrange equations obtained from its variation, a mathematical formulation of the electrostatic and elastic sub-problem, and FEM simulation details are given in **Appendix A** of **Supplementary Materials**. The ferroelectric, dielectric, and elastic properties of the BaTiO$_3$ [001] are listed in **Table AI** therein.

For the initial distribution of the polarization in the film, we used either a periodic 90-degree zigzag pattern of in-plane domain stripes or a regular arrangement of crossed 90-degree domain walls in the XZ-plane (see **Fig. A1a-c** in **Appendix A**). When we used a purely random noise as the initial distribution of polarization at room temperature, it relaxed to in-plane *a*-domain stripes without any flux-closure at the film surfaces. This circumstance can indicate that *a*-domain stripes are the energetic ground state configuration for a film in the rhombohedral phase. In fact, the energy density of the periodic stripe domains is -1.560 MPa, while the energy density of the vortex-antivortex pairs is -1.373 MPa (for the same polarity of antivortex cores) or -1.368 MPa (for the opposite polarity of antivortex cores) for a 10-nm BaTiO$_3$ film at room temperature.

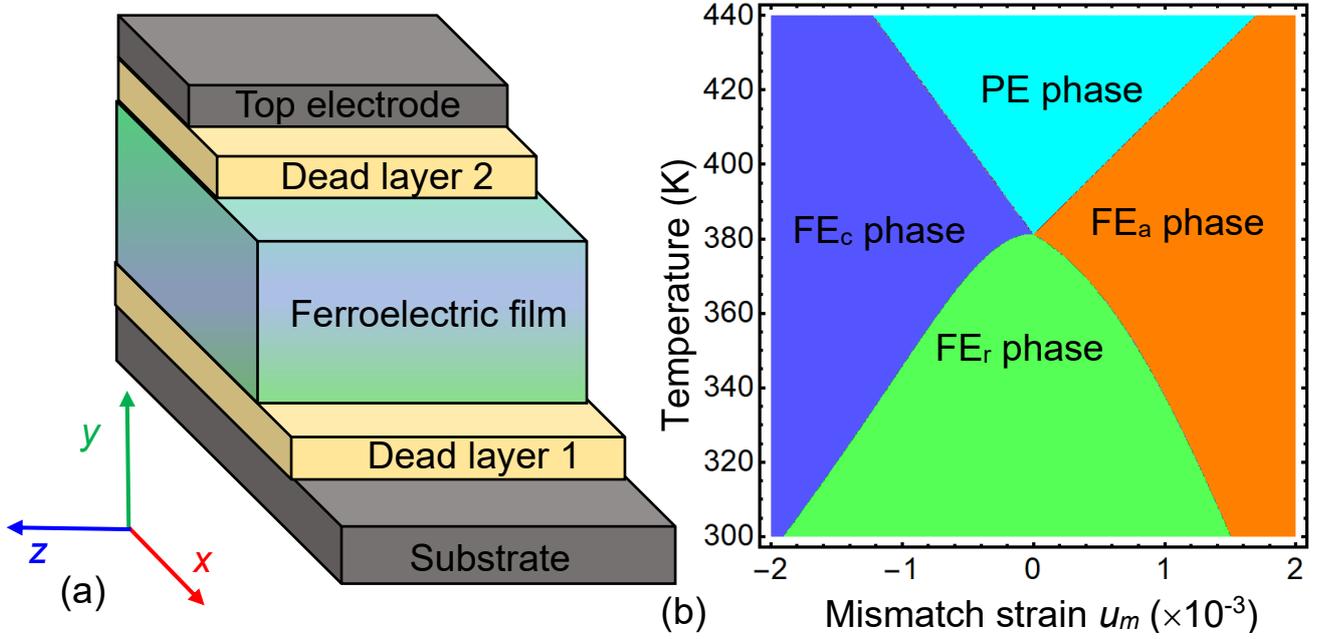

**FIGURE 1**. **(a)** The ferroelectric film placed between two paraelectric dead layers with a very high dielectric permittivity. Dead layers thickness is 0.8 nm and relative dielectric permittivity is 300. The mismatch strain exists at the interface "dead layer 1 – conducting substrate". The coordinate frame we are using is shown in the bottom left corner. **(b)** Phase diagram of a thick single-domain BaTiO$_3$ film on a rigid substrate. PE and FE denote the paraelectric and ferroelectric phases, respectively. Indices "c", "r" and "aa" corresponding to the tetragonal c-phase with an out-of-plane polarization, rhombohedral phase all three components of polarization, and a-phase with two in-plane components of polarization, respectively.

### III. Flexoelectric Coupling Influence on the Polarization Distribution

Our FEM studies revealed that the out-of-plane polarization component $P_y$ can be very sensitive to the flexoelectric coupling (shortly: "**flexocoupling**") for both stripe domains and vortices. However, the influence is rather different for each of them.

The influence of the flexocoupling on the in-plane periodic stripe domains in a 5-nm BaTiO$_3$ film is shown in **Fig. 2**. The dependence of the polarization magnitude $P = |\vec{P}|$ on $F_{ij}$ is negligible, and thus **Fig. 2a** is the same for positive, zero, and negative $F_{ij}$. This happens because the value of $P$ is governed by the in-plane polarization components, $P_x$ and $P_z$, and the amplitude of $P_x$ (about 20 μC/cm$^2$) is almost flexo-insensitive (see **Fig. 2e**). Although the flexocoupling modulates the amplitude of $P_z$, the modulation amplitude is very small (less than 0.6 μC/cm$^2$) in comparison with its almost constant value ~19 μC/cm$^2$ (see **Fig. 2g**). The flexocoupling significantly increases the amplitude of a small out-of-plane polarization component $P_y$ (from 0.1 μC/cm$^2$ to 0.4 μC/cm$^2$), and the "up" or "down" direction of $P_y$ is defined by the sign of flexocoupling coefficients $F_{ij}$. The amplitude of $P_y$ is proportional to the strength $|F_{ij}|$. This effect is illustrated qualitatively in **Fig. 2b, 2c** and **2d** plotted for negative, zero and positive $F_{ij}$, respectively. The quantitative influence of $F_{ij}$ on $P_y$ is clearly seen from z-profiles of $P_y$ in **Fig. 2f**, which are calculated at the film top surface for negative (blue curves), zero (black curves), and positive (red curves) flexoelectric coefficients $F_{ij}$. Note that the small value of $P_y$ at $F_{ij} = 0$ results from the electrostrictive coupling, and its phase depends on the initial conditions in the case.

Let us underline that the flexocoupling influence on $P_i$ revealed in this work and shown in **Fig. 2**, is in a qualitative agreement with the earlier predictions for the flexocoupling-induced Néel component of polarization at a "nominally uncharged" single domain wall in a bulk rhombohedral BaTiO3 [43] and BiFeO3 [41]. However, the flexocoupling influence on the periodic stripe *a*-domains in thin films is quantitatively different from that found in the case of single walls in a bulk rhombohedral material [41-43]. Using the results of Ref. [41], an approximate analytical expression for the Néel-type component

$$P_y \approx \frac{\varepsilon_0 \varepsilon_b f_Q}{1+2\beta\varepsilon_0\varepsilon_b} \frac{\partial P_x^2}{\partial z} + \frac{\varepsilon_0 \varepsilon_b q P_x}{1+2\beta\varepsilon_0\varepsilon_b}(P_s^2 - P_x^2) \qquad (1)$$

can be used as an estimate in the vicinity of the domain walls. Here the first term has flexoelectric nature, and the second term originates from electrostriction coupling. Constants $f_Q \approx F_{12}\frac{Q_{11}s_{12}-s_{11}Q_{12}}{s_{11}^2-s_{12}^2}$, $\beta \approx a_1 + (a_{12} + Q_{44}^2/2s_{44})P_s^2$, and $q \approx -\frac{Q_{11}s_{12}-s_{11}Q_{12}}{s_{11}^2-s_{12}^2}Q_{44}$, where $F_{12}$ is the flexoelectric coefficient, $Q_{ij}$ are electrostriction coefficients, $s_{12}$ are elastic compliances, $a_1$ and $a_{12}$ are LGD expansion coefficients, $P_s$ is the spontaneous polarization, $\varepsilon_b$ is a background permittivity [53], and $\varepsilon_0$ is the vacuum permittivity (see **Table AI**).

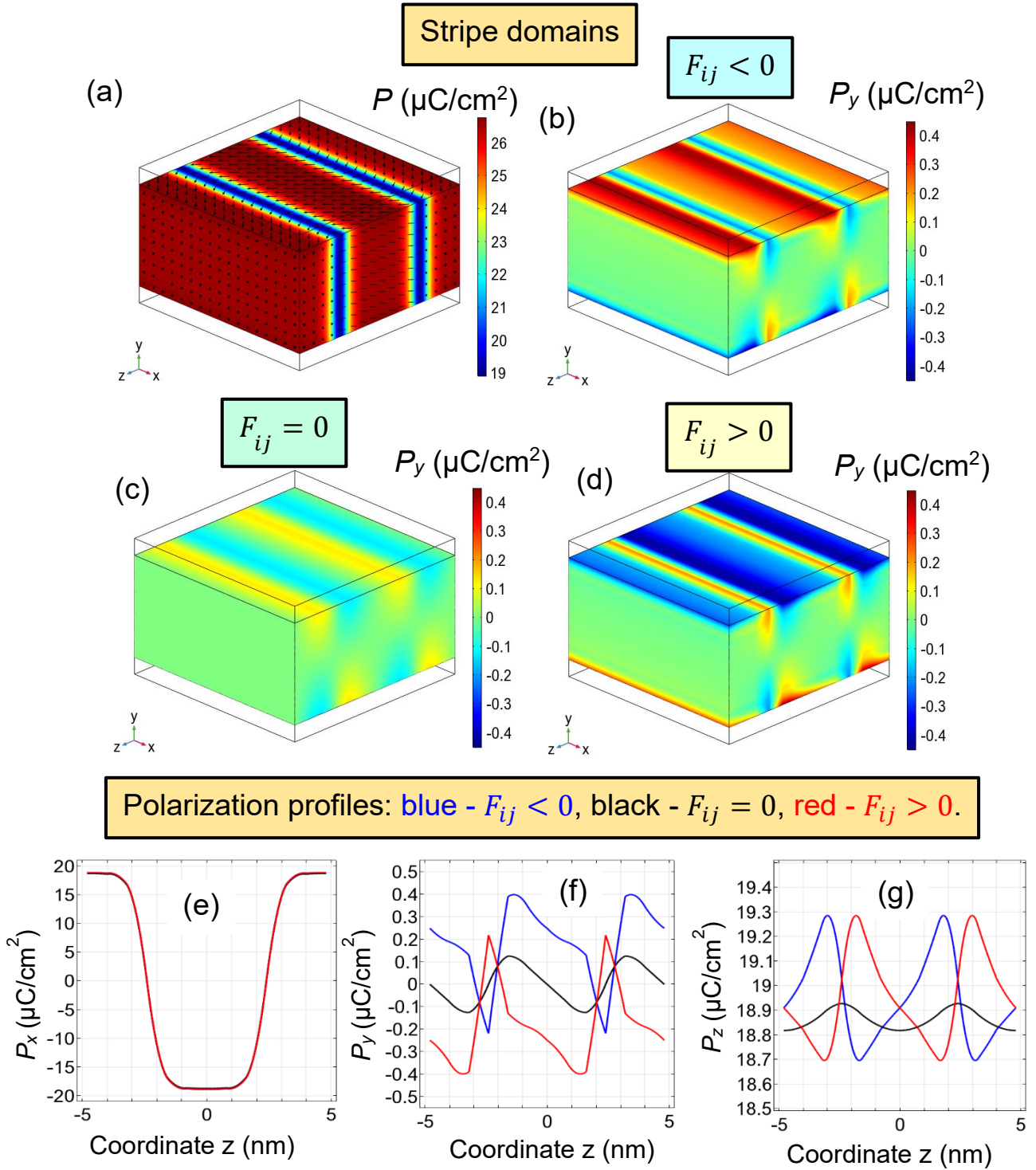

**FIGURE 2**. Regular stripe domains inside a thin strained BaTiO$_3$ film. Polarization magnitude $P$ (**a**) and out-of-plane component $P_y$ calculated for negative (**b**), zero (**c**) and positive (**d**) flexoelectric coefficients $F_{ij}$. (**e**)-(**g**) Polarization z-profiles calculated at the film top surface for negative (blue curves), zero (black curves) and positive (red curves) flexoelectric coefficients $F_{ij}$. The film thickness is 5 nm, mismatch strain $u_m$=0.2%, and the temperature $T$ is 300 K. The dependence of $P$ on $F_{ij}$ is negligible and not shown in the figure. Arrows in the plot (a) show the direction of polarization vector.

The influence of the flexocoupling on a periodic lattice of polarization vortices and antivortices is shown in **Figs. 3-5** and **Figs. B1-3** (in the **Supplement**) for a 9-nm BaTiO$_3$ film stretched with a misfit strain 0.1%. The computation cell, which translates periodically, is a rectangular parallelepiped consisting of two vortices separated by two antivortices (see **Fig. 3** and **Fig. B1**). The vortex and antivortex can be topologically distinguished by the winding number $W = \oint \frac{\alpha(\varphi)}{2\pi} dS$, that is the normalized line integral on a closed loop $S$ over the $\alpha(\varphi)$ that the in-plane component of the polarization $\vec{P}$ encloses with the *x*-axis [54]. The winding number counts the number of vortex ($W = +1$) and antivortex ($W = -1$) structures within the loop $S$ (see **Figs. 6a-c**).

Besides topological aspects, vortices and antivortices also differ concerning their energy. Although the correlation energy is similar in both cases, vortex structures are virtually divergence-free, such that they tend to form spontaneously in inhomogeneous structures to reduce the dipolar energy. By contrast, antivortices display strong divergences, leading to a distinct distribution of bound charges and a corresponding increase in dipolar energy. This behavior is known from magnetic structures, and the situation in ferroelectrics is analogous: here the divergence $div \vec{P}$ is relatively small in the vortex core case and quite significant in the antivortex one (see **Fig. A1d** in **Appendix A**). Because of these differences in energy, individual antivortices tend to dissolve in magnetic systems, and particular geometric shapes are required to stabilize them [55]. Antivortex structures typically occur for topological reasons as they form a link between two neighboring vortices with equal circulation. In magnetic structures, such situations are well-known in the form of "cross-tie" domain walls [56], which are essentially linear chains of alternating vortex-antivortex structures. However, we are not aware of a generalization of cross-tie domain wall structures to two-dimensional checkerboard-type vortex-antivortex arrays.

It appears that the dependence of the polarization magnitude $P$ on $F_{ij}$ is very weak, and so the images in **Fig. 3d** and **4c**, as well as *P*-profiles shown in **Figs. 5a** and **5b**, are almost $F_{ij}$-independent. In contrast to the magnitude, the distribution of the out-of-plane polarization component $P_y$ does depend on $F_{ij}$, i.e., it is "flexo-sensitive", but the sensibility is only significant for a vortex region (as will be shown below).

The antivortex has a wide and smooth dipolar core, whose rounded rectangular-like shape and other features are weakly sensitive to the flexocoupling (see **Figs. 3a-c** and **Fig. B2a-c**). The direction of the component $P_y$ in the antivortex core is defined by initial conditions (compare **Fig. 4a** and **4b**), but not by the $F_{ij}$ structure. The magnitude of $P_y$ in the antivortex core is relatively high, reaching (10 – 15) μC/cm$^2$ for thin films (see **Fig. 5c,e,g**).

The vortex has a very prolate asymmetrical spike-like quadrupolar core, whose polarity is defined by the sign of $P_y$, is controlled by the flexocoupling (compare blue and reddish vortex cores in **Figs. 3a-** **c** and the quadrupolar areas inside dotted ellipses in **Figs. 4d-f**). The sign of $F_{ij}$ determines the sign of the polarization and its value in the vortex core, as it is seen from the color images in **Figs. 4d-f**, and especially from the *y*-profiles in **Fig. 5d, 5f** and **5h** calculated for $F_{ij} < 0$, $F_{ij} = 0$ and $F_{ij} > 0$, respectively. The vortex core polarization becomes significantly smaller for $F_{ij} = 0$ (see **Fig. 5f**). At the nominal values of $F_{ij}$ (listed in **Table AI**) the maximal polarization in the vortex core reaches (1 – 1.5) µC/cm², depending on the film surface (see **Fig. 5d, 5f** and **5h**). An axial asymmetry of the quadrupolar core is due to different elastic boundary conditions at the mechanically free top surface, where the normal stress is absent, and at the bottom surface clamped to a rigid substrate with a mismatch strain. The asymmetry is clearly seen by comparing the small and high-contrast reddish part with the smooth and more extended blue part of the quadrupolar core in **Figs. 4d** and **4f**, as well as from the comparison of solid and dashed curves in **Figs. 5d** and **5h**. Notably that the transformation $F_{ij} \rightarrow -F_{ij}$ inverts the vortex core ($P_y \rightarrow -P_y$) with respect to the y-axis.

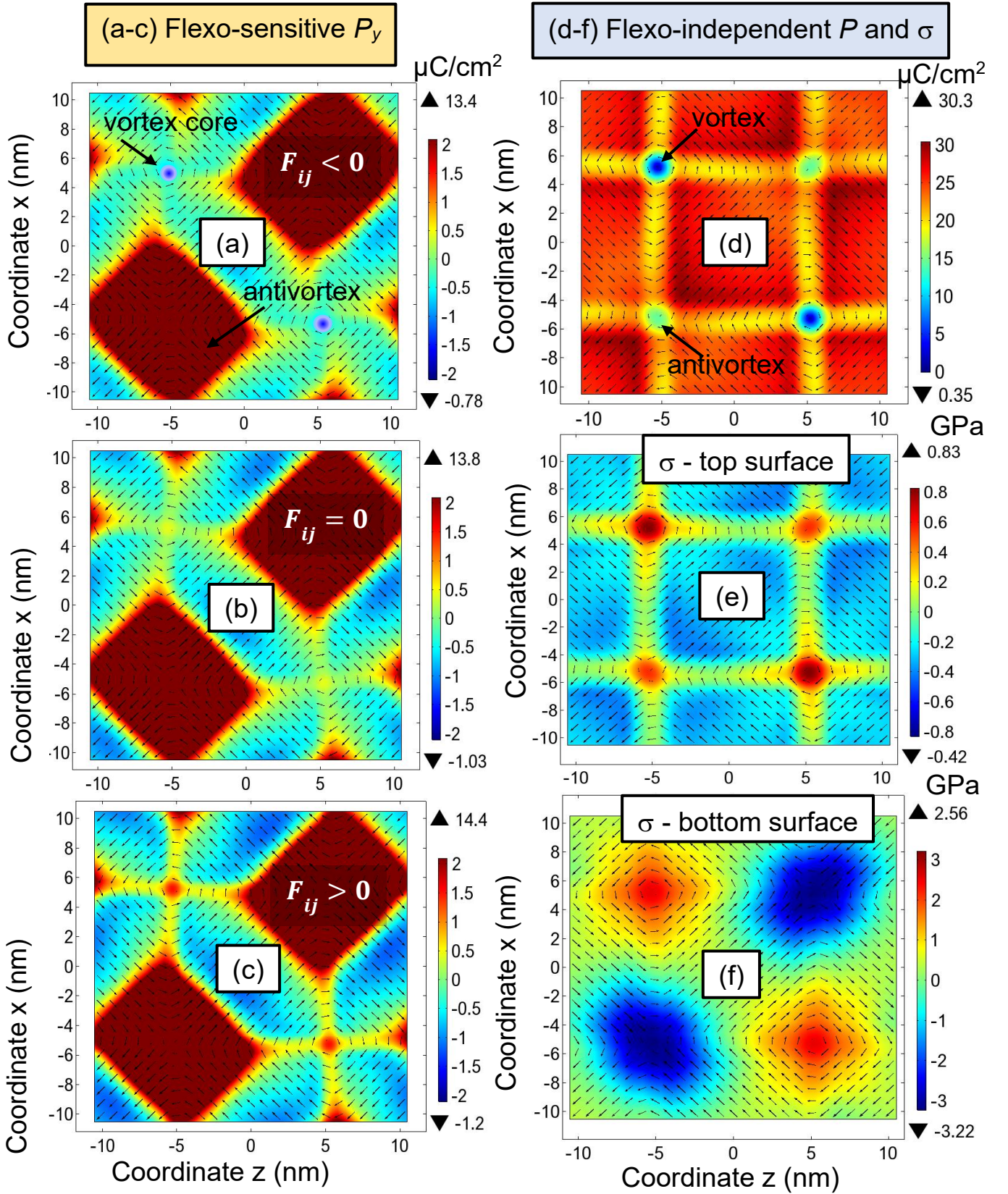

**FIGURE 3**. The XZ-sections of the polarization out-of-plane component $P_y$ at the film top surface calculated for negative (**a**), zero (**b**) and positive (**c**) flexoelectric coefficients $F_{ij}$. The XZ-sections of the polarization magnitude $P$ (**d**) and hydrostatic stress $\sigma$ (**e, f**), which are almost flexo-independent. Arrows show polarization direction in the vortex-antivortex region. The thickness of BaTiO$_3$ film is 9 nm, mismatch strain $u_m$=0.1%, and $T$ = 300 K.

In general, the difference in the behavior of the antivortex and vortex cores (shown in **Figs. 3a-c** and **Fig. 4**) can be explained by the difference of elastic fields in the regions. The hydrostatic stress, $\sigma = \sigma_{11} + \sigma_{22} + \sigma_{33}$, shown in **Fig. 3e** (at the film top surface) and **3f** (at the film bottom surface), and in **Fig. B3** (at both surfaces), is well-localized in the antivortex and vortex regions. The stress is maximal and positive in the vortex core at the film top surface, where it is significantly smaller than the stress in the antivortex core. The situation is opposite at the film bottom surface, where $\sigma$ is maximal and negative in the antivortex region, and it becomes positive and significantly smaller in magnitude in the vortex region. The spatial distribution of a dilatational strain, $u = u_{11} + u_{22} + u_{33}$, is very similar to the stress, and thus not shown here. Although the components of elastic stress and strain are virtually independent on the $F_{ij}$ value, their gradients convoluted with flexoelectric coefficients lead to the flexo-sensitivity of the vortices.

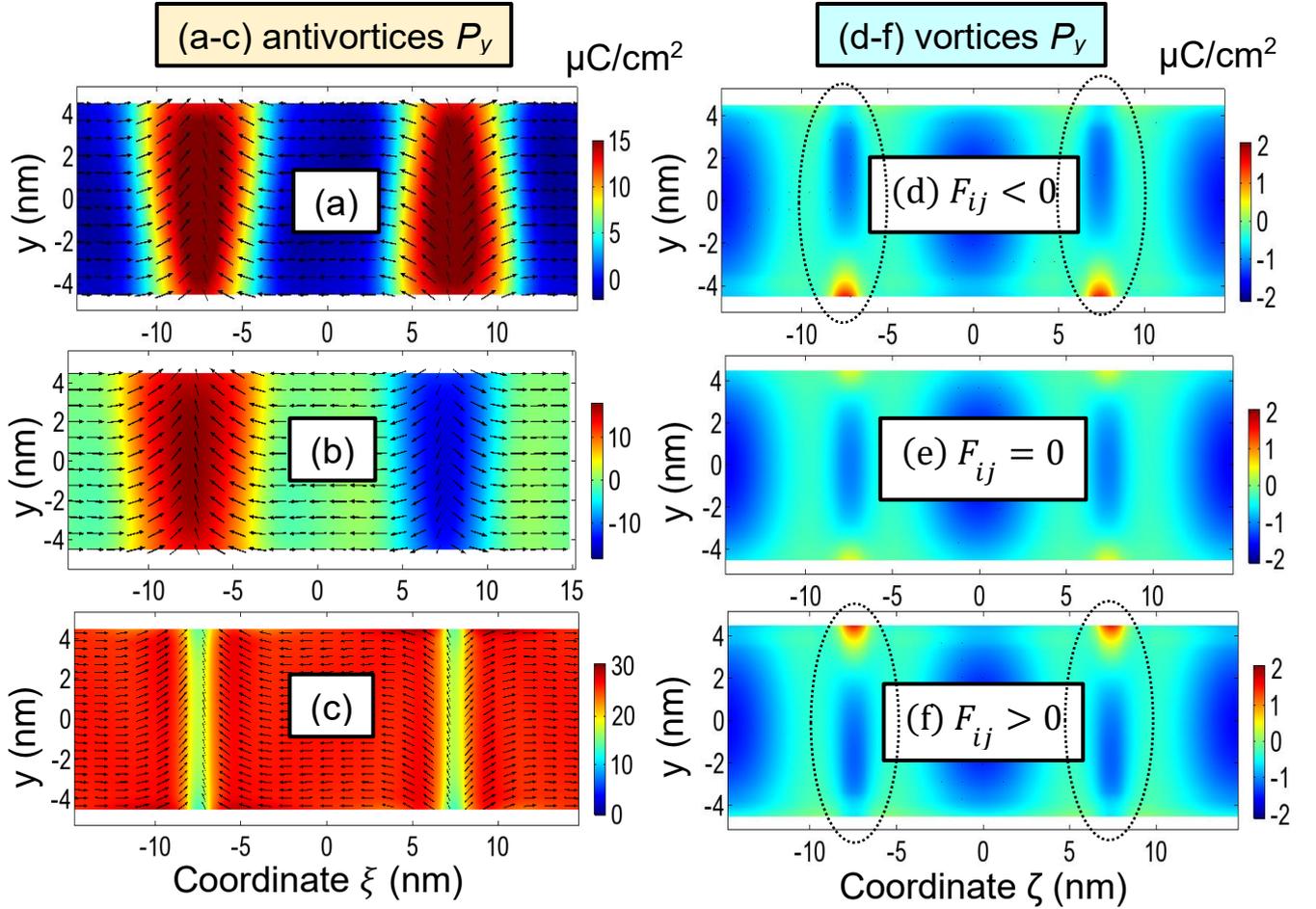

**FIGURE 4**. The $\xi Y$-sections of the out-of-plane polarization component $P_y$ for different seedings **(a-b)** and the magnitude $P$ **(c)**, which are almost flexo-independent. Flexo-sensitive $\zeta Y$-sections of the $P_y$ calculated for negative **(d)**, zero **(e)** and positive **(f)** flexoelectric coefficients $F_{ij}$. Coordinates $\xi = (x+z)/\sqrt{2}$ and $\zeta = (x-z)/\sqrt{2}$. Arrows in the plots (a)-(c) show polarization direction. The thickness of BaTiO$_3$ film is 9 nm, mismatch strain $u_m$=0.1%, and $T = 300$ K.

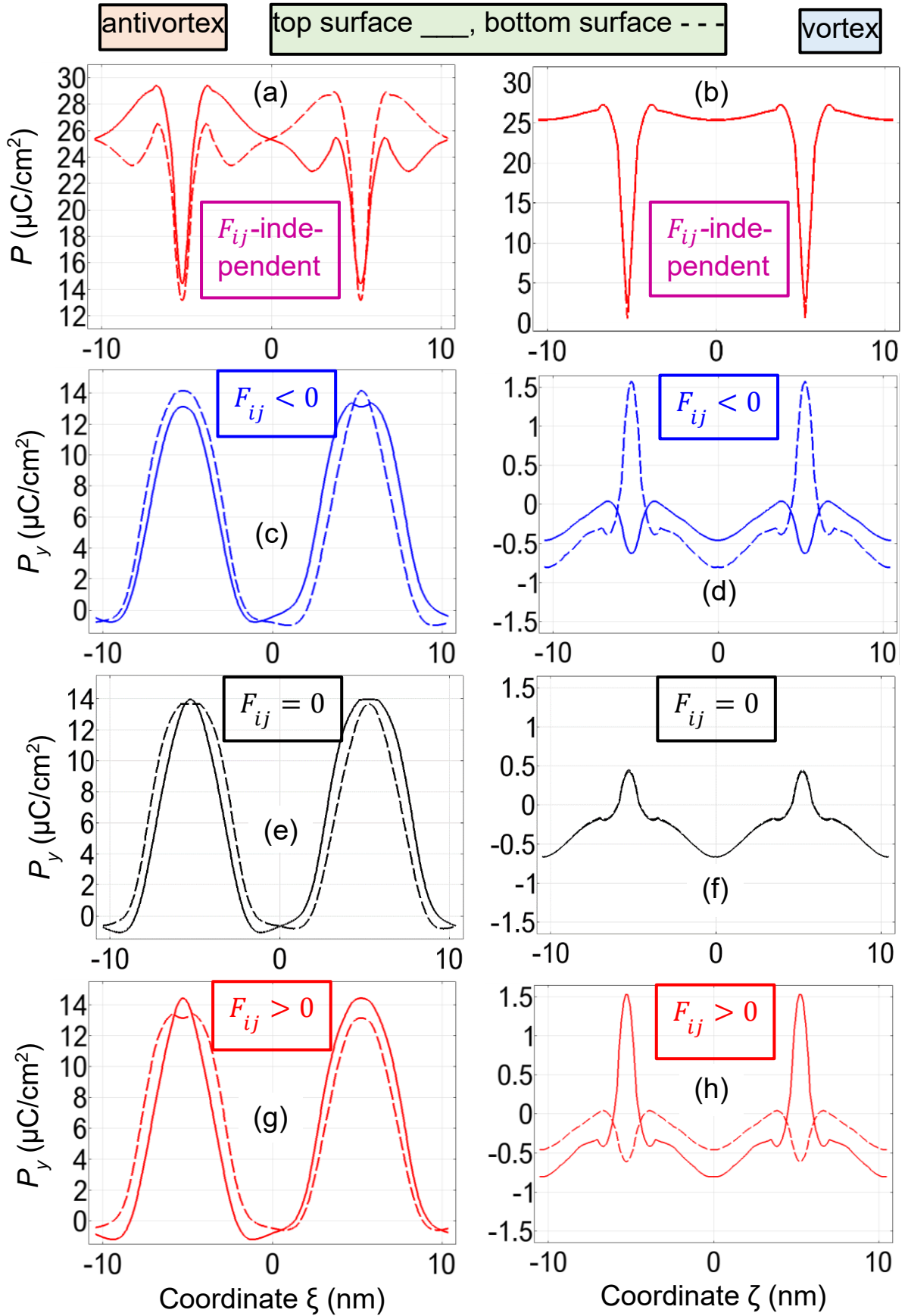

**FIGURE 5**. Profiles of the flexo-independent antivortices (left column) and flexo-sensitive vortices (right column) at the top (solid curves) and bottom (dashed curves) surfaces of a thin strained BaTiO$_3$ film. **(a)-(b)** Flexo-independent $\xi$- and $\zeta$-profiles of the polarization magnitude $P$. **(c)-(h)** Flexo-sensitive $\xi$- and $\zeta$-profiles of the out-

of-plane polarization component $P_y$ for negative (blue curves), zero (black curves) and positive (red curves) flexoelectric coefficients $F_{ij}$. Coordinates $\xi = (x+z)/\sqrt{2}$ and $\zeta = (x-z)/\sqrt{2}$. The film thickness is 9 nm, mismatch strain $u_m$=0.1%, and $T = 300$ K.

The behavior of the flexo-sensitive polarization vortex is qualitatively the same as the flexon-type polarization in cylindrical nanoparticles [51]. We argue that the flexo-sensitive vortex cores can be interpreted as a manifestation of "flexons" [51] in thin films. To ascertain this conjecture, a topological index of these structures is calculated and analyzed below.

The topological index $n(y) = \frac{1}{4\pi}\int_S \vec{p}\left[\frac{\partial \vec{p}}{\partial x} \times \frac{\partial \vec{p}}{\partial z}\right]dxdz$ [57], where $\vec{p} = \frac{\vec{P}}{P}$ is the unit polarization orientation and the integration is performed over the vortex cross-section $S = \{x,z\}$, quantifies the chirality of the polarization structure. Using the method explained in Appendix D of Ref.[51], the approximate y-dependence of the topological index is

$$n(y) \approx -\frac{P_y(x=0,y,z=0)}{2P(x=0,y,z=0)}. \tag{2}$$

Similarly to the flexon discussed in Ref.[51], $n(y)$ is a normalized profile of the out-of-plane polarization component, and the coordinate origin is placed in the vortex axis. Note that the expression (2) is an approximation since the distribution of $P_y(x,y,z)$ is not fully axially symmetric with respect to the vortex axis. The axial asymmetry appears in the form of a square mesh between the vortices (see e.g., **Fig. 3**). Although $P_y$ is nonzero for $F_{ij} = 0$ due to the electrostriction coupling, two Bloch points ($P = 0$) located symmetrically under the film surfaces exist for that case (see **Fig. 6d**). The depth profile of $P_y$ becomes asymmetric with respect to the film surfaces for $F_{ij} \neq 0$, and only one Bloch point exists in this case (see **Fig. 6e**). The transformation $F_{ij} \rightarrow -F_{ij}$ leads to the transformation $P_y \rightarrow -P_y$ (compare red and blue curves in **Fig. 6e**). Hence, the sign of $P_y$, and so the sign of $n(y)$, is defined by the sign of $F_{ij}$, since the flexo-induced contribution to the polarization component $P_y$ significantly dominates over the electrostriction contribution. The result corroborates the "flexon" character [51] of the vortex polarization. The dependence $n(y)$ is shown in **Fig. 6e** for zero, positive and negative $F_{ij}$, respectively. Since the value $P(0,y,0)$ coincides with $|P_y(0,y,0)|$ at the vortex axis, and $P_y(0,y,0) = P(0,y,0) = 0$ in the Bloch point, the topological index jumps in that point from -½ to +½ (red curve in **Fig. 6e**), or from +½ to -½ (blue curve in **Fig. 6e**), depending on the $F_{ij}$ sign. In any case, $n(y) = \pm 1/2$ at the film surfaces. The result demonstrates the surface localization of the flexo-sensitive vortices, similarly to the "edge" localization of flexons in cylindrical nanoparticles [51]. The topological index, which can be interpreted as the degree to which a structure is chiral, changes sign from one surface to the other, and changes sign upon reversal of the sign of $F_{ij}$. Since $|P_y| \sim |F_{12}|$, the index $n(y)$ increases in magnitude with increasing absolute value of $|F_{ij}|$. These properties make obvious the clear correlation between the flexocoupling and the formation of chiral polar vortices in thin ferroelectric films.

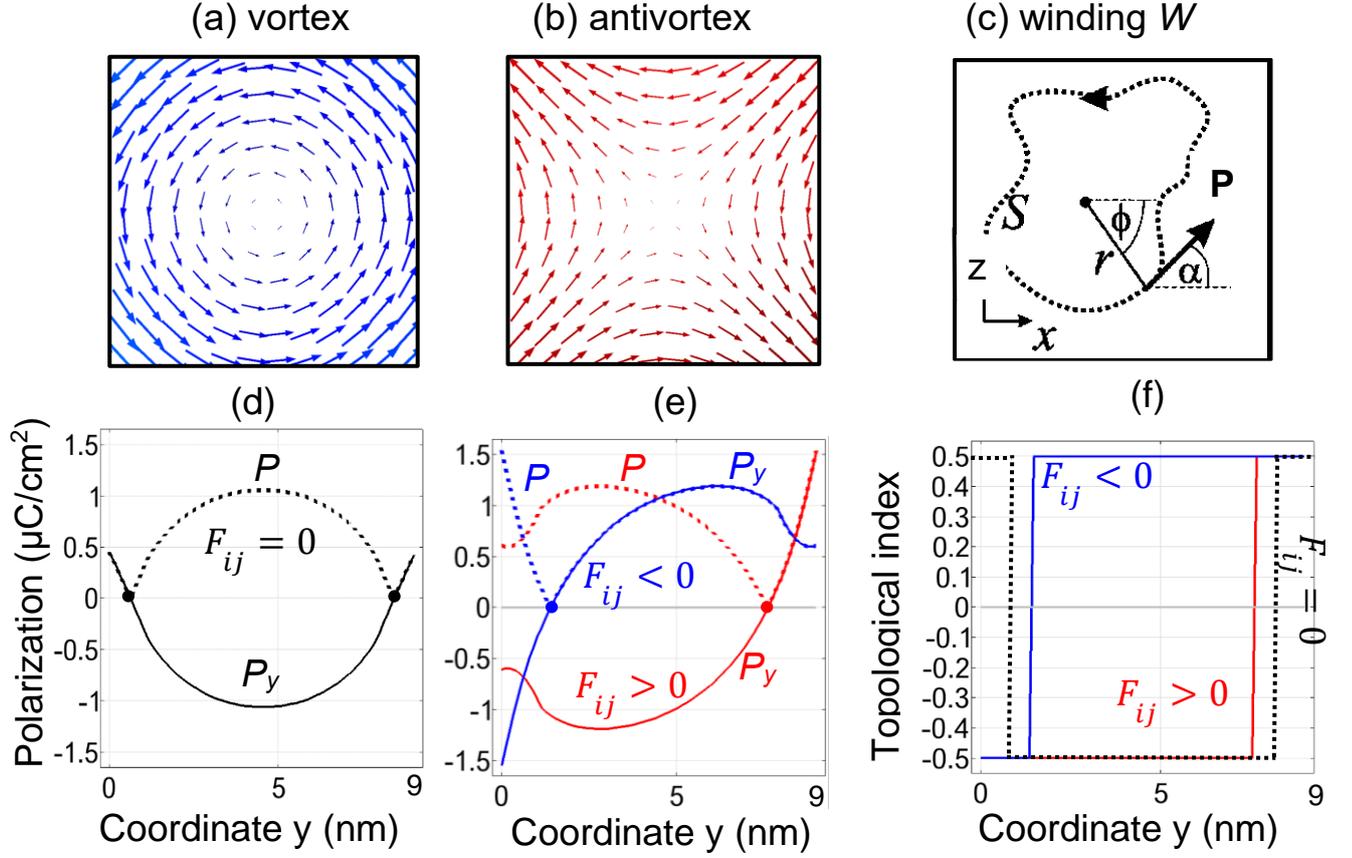

**FIGURE 6.** (a) A ferroelectric vortex ($W = +1$) and (b) antivortex ($W = -1$), whose topology can be distinguished by the winding number $W$ (c). Part (c) is redrawn from Ref. [54]. (d-e) Depth profiles of the out-of-plane polarization component $P_y$ (solid curves) and magnitude $P$ (dotted curves) calculated at the vortex axis for zero (black curves, d), negative (blue curves, e) and positive (red curves, e) flexoelectric coefficients $F_{ij}$. Circles are Bloch points, where $P = 0$. (f) The y-profile of the polarization's topological index $n(y)$ for zero (black dotted lines), positive (solid red lines), and negative (solid blue lines) $F_{ij}$. The film thickness is 9 nm, mismatch strain $u_m$=0.1%, and $T = 300$ K.

To the best of our understanding, the experiment [46] indicates the possible existence of flexo-sensitive vortices. However, most phase-field simulations (see e.g., [58]) either did not include the flexoelectric effect, or they reveal other trends [49]. Indeed, the flexo-sensitive vortices are "hidden" by flexo-insensitive antivortices, whose out-of-plane polarization is much higher. Moreover, when we increase the transverse size of the computational cell, the core of the antivortex becomes a complex structure resembling an r-phase domain. Still, the polarization remains strictly vertical in the core region, as for the c-phase domain.

To define the temperature interval, where the flexo-sensitive stripes or vortices are stable or meta-stable, we performed FEM studies of the polarization distribution in the film in the temperature range from 200 K to 440 K. A stripe-domain configuration with a flexo-sensitive out-of-plane polarization component $P_y$ is stable at temperatures lower than 415 K; at higher temperatures it becomes faint as the film transforms to the paraelectric phase.

A vortex-like configuration with a flexo-sensitive out-of-plane polarization component $P_y$ is metastable at temperatures lower than 410 K (we use the word "metastable" here to underline that the domain stripes have lower energy). The flexo-insensitive antivortex core disappears above 320 K. The dipolar core of the vortex gradually disappears above 415 K, when the vortex-antivortex structure transforms into a circular vortex-antivortex structure without both cores, which exists up to the paraelectric transition at about 420 K. The relatively wide temperature range ($200 < T < 410$ K) of the flexo-sensitive vortices meta-stability gives us the hope that the domain morphology can be observed experimentally in thin ferroelectric films by PFM, cAFM [45-47] and nonlinear optical microscopy [48] methods.

The disappearance of antivortex cores when the temperature is decreased below (315 – 320) K is related to the transition from the monoclinic r-phase (with all three polarization components) to the orthorhombic a-phase (with two in-plane components of polarization). The monoclinic-orthorhombic transition occurs at about 340 K for thick films; and its temperature decrease to 320 K in thin films with dead layers is caused by the depolarization field arising from the layers.

## IV. Conclusion

In this theoretical work, we have considered the influence of the flexocoupling on the spatial distribution and the temperature behavior of the ferroelectric polarization for different types of stable domain structures in thin ferroelectric films, such as periodic stripe domains and arrays of vortex-antivortex pairs. Our FEM simulations and analytical calculations reveal that an out-of-plane polarization component can be very sensitive to the flexocoupling for both stripe domains and vortices. However, the influence is rather different. Namely, the flexocoupling significantly increases the amplitude of a small out-of-plane polarization component in the stripe domains. The "up" or "down" direction of this out-of-plane component depends on the sign of the flexocoupling coefficients. The vortex has a spike-like quadrupolar core controlled by the flexocoupling, whereas the antivortex has a wide smooth dipolar core, whose shape and other features are weakly insensitive to the coupling. The origin of the flexo-sensitive vortex is due to the system's tendency to minimize its elastic and electrostatic energy, because a vortex domain wall (in fact, a flux-closure domain) creates a much weaker depolarization field compared to the field of a charged domain wall.

We interpret the flexo-sensitivity of the vortex cores as a variant of the effects that stabilize the recently revealed "flexons" [51] in thin films. To the best of our knowledge, such flexo-sensitive vortices have not been reported previously, while several experiments [46, 49] suggest their existence.

The relatively wide temperature range ($200 < T < 410$ K) in which the flexo-sensitive vortices are meta-stable gives us hope that these structures can be observed experimentally in thin ferroelectric films by PFM, cAFM, and nonlinear optical microscopy methods. However, in thin films, the PFM observation of the flexo-sensitive vortices is complicated by the presence of much higher contrast antivortices, which are flexo-insensitive.

**Acknowledgements.** This material is based upon work (S.V.K.) supported by the U.S. Department of Energy, Office of Science, Office of Basic Energy Sciences, and performed at the Center for Nanophase Materials Sciences, a US Department of Energy Office of Science User Facility. A portion of FEM was conducted at the Center for Nanophase Materials Sciences, which is a DOE Office of Science User Facility (CNMS Proposal ID: 257). A.N.M. work is supported by the National Academy of Sciences of Ukraine (the Target Program of Basic Research of the National Academy of Sciences of Ukraine "Prospective basic research and innovative development of nanomaterials and nanotechnologies for 2020 - 2024", Project № 1/20-Н, state registration number: 0120U102306). R.H. acknowledges funding from the French National Research Agency through contract ANR-18-CE92-0052 "TOPELEC".

# Supplementary Materials
## APPENDIX A. Mathematical Formulation of the Problem and FEM Details
## A. Mathematical Formulation of the Problem

The LGD free energy functional $G$ additively includes a Landau expansion on powers of 2-4-6 of the polarization ($P_i$), $G_{Landau}$; a polarization gradient energy contribution, $G_{grad}$; an electrostatic contribution, $G_{el}$; the elastic, electrostriction, flexoelectric contributions, $G_{es+flexo}$; and a surface energy, $G_S$. It has the form:

$$G = G_{Landau} + G_{grad} + G_{el} + G_{es+flexo} + G_S, \qquad (A.1a)$$

$$G_{Landau} = \int_{V_f} d^3r \left[ a_i P_i^2 + a_{ij} P_i^2 P_j^2 + a_{ijk} P_i^2 P_j^2 P_k^2 \right], \qquad (A.1b)$$

$$G_{grad} = \int_{V_f} d^3r \frac{g_{ijkl}}{2} \frac{\partial P_i}{\partial x_j} \frac{\partial P_k}{\partial x_l}, \qquad (A.1c)$$

$$G_{el} = -\int_{V_f} d^3r \left( P_i E_i + \frac{\varepsilon_0 \varepsilon_b}{2} E_i E_i \right) - \frac{\varepsilon_0}{2} \int_{V_d} \varepsilon_{ij}^d E_i E_j d^3r, \qquad (A.1d)$$

$$G_{es+flexo} = -\int_{V_f} d^3r \left[ \frac{s_{ijkl}}{2} \sigma_{ij} \sigma_{kl} + Q_{ijkl} \sigma_{ij} P_k P_l + F_{ijkl} \left( \sigma_{ij} \frac{\partial P_k}{\partial x_l} - P_k \frac{\partial \sigma_{ij}}{\partial x_l} \right) \right] \qquad (A.1e)$$

$$G_S = \frac{1}{2} \int_S d^2r\, a_{ij}^{(S)} P_i P_j. \qquad (A.1f)$$

Here $V_f$ and $V_d$ are the film and dead layer volume, respectively. The coefficient $a_i$ linearly depends on temperature $T$, $a_i(T) = \alpha_T[T - T_C]$, where $\alpha_T$ is the inverse Curie-Weiss constant and $T_C$ is the ferroelectric Curie temperature renormalized by electrostriction and surface tension. Tensor components $a_{ij}$ are regarded as temperature-independent. The tensor $a_{ij}$ is negatively defined for the BaTiO$_3$ undergoing the first order transition to the paraelectric phase. The higher nonlinear tensor $a_{ijk}$ and the gradient coefficients tensor $g_{ijkl}$ are positively defined and regarded as temperature-independent. The following designations are used in Eq.(A.1e): $\sigma_{ij}$ is the stress tensor, $s_{ijkl}$ is the elastic compliances tensor, $Q_{ijkl}$ is the electrostriction tensor, and $F_{ijkl}$ is the flexoelectric tensor.

Allowing for the Khalatnikov mechanism of polarization relaxation [59], minimization of the free energy (A.1) with respect to polarization leads to three coupled time-dependent Euler-Lagrange equations for polarization components inside the film, $\frac{\delta G}{\delta P_i} = -\Gamma \frac{\partial P_i}{\partial t}$, where $i = 1, 2, 3$. The explicit form of the equations for a ferroelectric crystal with m3m parent symmetry is:

$$\Gamma \frac{\partial P_1}{\partial t} + 2P_1(a_1 - Q_{12}(\sigma_{22} + \sigma_{33}) - Q_{11}\sigma_{11}) - Q_{44}(\sigma_{12}P_2 + \sigma_{13}P_3) + 4a_{11}P_1^3 + 2a_{12}P_1(P_2^2 + P_3^2) +$$

$$6a_{111}P_1^5 + 2a_{112}P_1(P_2^4 + 2P_1^2 P_2^2 + P_3^4 + 2P_1^2 P_3^2) + 2a_{123}P_1 P_2^2 P_3^2 - g_{11}\frac{\partial^2 P_1}{\partial x_1^2} - g_{44}\left(\frac{\partial^2 P_1}{\partial x_2^2} + \frac{\partial^2 P_1}{\partial x_3^2}\right) =$$

$$-F_{11}\frac{\partial \sigma_{11}}{\partial x_1} - F_{12}\left(\frac{\partial \sigma_{22}}{\partial x_1} + \frac{\partial \sigma_{33}}{\partial x_1}\right) - F_{44}\left(\frac{\partial \sigma_{12}}{\partial x_2} + \frac{\partial \sigma_{13}}{\partial x_3}\right) + E_1$$

$$(A.2a)$$

$$\Gamma \frac{\partial P_2}{\partial t} + 2P_2(a_1 - Q_{12}(\sigma_{11} + \sigma_{33}) - Q_{11}\sigma_{22}) - Q_{44}(\sigma_{12}P_1 + \sigma_{23}P_3) + 4a_{11}P_2^3 + 2a_{12}P_2(P_1^2 + P_3^2) +$$

$$6a_{111}P_2^5 + 2a_{112}P_2(P_1^4 + 2P_2^2P_1^2 + P_3^4 + 2P_2^2P_3^2) + 2a_{123}P_2P_1^2P_3^2 - g_{11}\frac{\partial^2 P_2}{\partial x_2^2} - g_{44}\left(\frac{\partial^2 P_2}{\partial x_1^2} + \frac{\partial^2 P_2}{\partial x_3^2}\right) =$$

$$-F_{11}\frac{\partial \sigma_{22}}{\partial x_2} - F_{12}\left(\frac{\partial \sigma_{11}}{\partial x_2} + \frac{\partial \sigma_{33}}{\partial x_2}\right) - F_{44}\left(\frac{\partial \sigma_{12}}{\partial x_1} + \frac{\partial \sigma_{23}}{\partial x_3}\right) + E_2$$

(A.2b)

$$\Gamma \frac{\partial P_3}{\partial t} + 2P_3(a_1 - Q_{12}(\sigma_{11} + \sigma_{22}) - Q_{11}\sigma_{33}) - Q_{44}(\sigma_{13}P_1 + \sigma_{23}P_2) + 4a_{11}P_3^3 + 2a_{12}P_3(P_1^2 + P_2^2) +$$

$$6a_{111}P_3^5 + 2a_{112}P_3(P_1^4 + 2P_3^2P_1^2 + P_2^4 + 2P_2^2P_3^2) + 2a_{123}P_3P_1^2P_2^2 - g_{11}\frac{\partial^2 P_3}{\partial x_3^2} - g_{44}\left(\frac{\partial^2 P_3}{\partial x_1^2} + \frac{\partial^2 P_3}{\partial x_2^2}\right) =$$

$$-F_{11}\frac{\partial \sigma_{33}}{\partial x_3} - F_{12}\left(\frac{\partial \sigma_{11}}{\partial x_3} + \frac{\partial \sigma_{22}}{\partial x_3}\right) - F_{44}\left(\frac{\partial \sigma_{13}}{\partial x_1} + \frac{\partial \sigma_{23}}{\partial x_2}\right) + E_3$$

(A.2c)

The temperature-dependent Khalatnikov coefficient $\Gamma$ [60] determines the relaxation time of the polarization $\tau_K = \Gamma/|\alpha|$. Consequently, $\tau_K$ typically varies in the range $(10^{-9} - 10^{-6})$ seconds for temperatures far from $T_C$. As argued by Hlinka et al. [61], we assumed that $g'_{44} = -g_{12}$ in Eqs.(A.2). The boundary condition for polarization at the film-dead layer interface accounts for the flexoelectric effect:

$$a_{ij}^{(S)}P_j + \left(g_{ijkl}\frac{\partial P_k}{\partial x_l} - F_{klij}\sigma_{kl}\right)n_j\bigg|_{x_3=0,h} = 0 \tag{A.3}$$

where **n** is the outer normal to the surface, $i = 1, 2, 3$. In our FEM studies, we use the so-called "natural" boundary conditions corresponding to $a_{ij}^{(S)} = 0$. To model the film infinity in the lateral directions "1" and "2", The periodic boundary conditions are valid at the sidewalls of the computation region.

The electric field components $E_i$ in Eqs.(A.2) are derived from the electric potential $\varphi$ in a conventional way, $E_i = -\partial\varphi/\partial x_i$. The potential $\varphi_f$ satisfies the Poisson equation in the ferroelectric film (subscript "$f$"):

$$\varepsilon_0\varepsilon_b\left(\frac{\partial^2}{\partial x_1^2} + \frac{\partial^2}{\partial x_2^2} + \frac{\partial^2}{\partial x_3^2}\right)\varphi_f = \frac{\partial P_i}{\partial x_i}, \qquad 0 \leq x_3 \leq h. \tag{A.4a}$$

The electric potential $\varphi_d$ in the dead layer satisfies the Laplace equation (subscript "$d$"):

$$\varepsilon_0\varepsilon_e\left(\frac{\partial^2}{\partial x_1^2} + \frac{\partial^2}{\partial x_2^2} + \frac{\partial^2}{\partial x_3^2}\right)\varphi_d = 0, \qquad -d < x_3 < 0 \cup h < x_3 < h+d. \tag{A.4b}$$

Equations (A.4) are supplemented with the continuity conditions for electric potential and the normal components of the electric displacements at the film surfaces:

$$(\varphi_d - \varphi_f)\big|_{x_3=0,h} = 0, \quad \mathbf{n}(\mathbf{D}_d - \mathbf{D}_f)\big|_{x_3=0,h} = 0. \tag{A.4a}$$

The voltage is fixed at the electrodes:

$$\varphi_d|_{x_3=-d} = 0, \quad \varphi_d|_{x_3=h+d} = U. \tag{A.4c}$$

and periodic boundary conditions for electric potential are valid at the sidewalls of the computation region.

Elastic equations of state follow from the variation of the energy (A.1e) with respect to elastic stress, $\frac{\delta G}{\delta \sigma_{ij}} = -u_{ij}$:

$$s_{ijkl}\sigma_{kl} + Q_{ijkl}P_k P_l + F_{ijkl}\frac{\partial P_l}{\partial x_k} = u_{ij}, \quad 0 \leq x_3 \leq h, \tag{A.5a}$$

$$\sigma_{ij} = c_{ijkl}u_{kl} - q_{ijkl}P_k P_l - f_{ijkl}\frac{\partial P_l}{\partial x_k}, \quad 0 \leq x_3 \leq h, \tag{A.5b}$$

where $u_{ij}$ is the strain tensor components related to the displacement components $U_i$ in the following way: $u_{ij} = (\partial U_i/\partial x_j + \partial U_j/\partial x_i)/2$.

Equations (A.5) should be considered along with equations of mechanical equilibrium

$$\partial \sigma_{ij}(\boldsymbol{x})/\partial x_i = 0, \tag{A.6}$$

compatibility equations, $e_{ikl}e_{jmn}\partial^2 u_{ln}(\boldsymbol{x})/\partial x_k \partial x_m = 0$, which are equivalent to the mechanical displacement vector $U_i$ continuity [62]. The boundary conditions for elastic stresses and displacements at the film surfaces are conventional continuity of elastic displacement vector and normal stress components:

$$\left(U_i^d - U_i^f\right)\Big|_{x_3=0,h} = 0, \quad \left(\sigma_{3j}^d - \sigma_{3j}^f\right)\Big|_{x_3=0,h} = 0. \tag{A.7a}$$

The boundary conditions for elastic stresses and displacements at the dead layer surfaces account for the free top surface and mismatch strain at the substrate electrode ($u_{11} = u_{22} = u_m$):

$$\sigma_{3j}^d\Big|_{x_3=h+d} = 0, \quad U_3^d\Big|_{x_3=-d} = 0, \quad U_1^d\Big|_{x_3=-d} = x_1 u_m, \quad U_2^d\Big|_{x_3=-d} = x_2 u_m. \tag{A.7b}$$

### B. Finite Element Modelling and Analytical Calculations Details

FEM simulations are performed in COMSOL@MultiPhysics software, using electrostatics, solid mechanics, and general math (PDE toolbox) modules. The size of the computational region is not less than 40×40×160 nm³, and is commensurate with the cubic unit cell constant (about 0.4 nm) of BaTiO$_3$ at room temperature. The minimal size of a tetrahedral element in a mesh with fine discretization is equal to the unit cell size, 0.4 nm, the maximal size is 1.6 nm. The dependence on the mesh size is verified by increasing the minimal size to 0.8 nm. We verified that this only results in minor changes in the electric polarization, electric field, and elastic stress and strain, such that the spatial distribution of each of these quantities becomes less smooth (i.e., they contain numerical errors in the form of a small random noise). However, when using these larger cell sizes, all significant details remain visible, and more importantly, the system energy remains essentially the same with an accuracy of about 0.1%.

LGD coefficients and other material parameters of BaTiO$_3$ are listed in **Table AI.** Polarization components in the initial distribution, whose relaxation leads to the lateral grid of vortices and antivortices, are shown in **Fig. A1.** Note that we superimposed a slight random noise to prevent the formation of unstable equilibrium states arising from idealized initial conditions.

**Table AI.** LGD coefficients and other material parameters of BaTiO$_3$

| Coefficient | Numerical value |
|---|---|
| $\varepsilon_{b,\,e}$ | $\varepsilon_b = 7$ (core background)    $\varepsilon_e = 10$ (surrounding) |
| $a_i$   (in mJ/C$^2$) | $a_1 = 3.34(T-381)\cdot10^5$,   $\alpha_T = 3.34\cdot10^5$         ($a_1 = -2.94\cdot10^7$ at 298 K) |
| $a_{ij}$   (in m$^5$J/C$^4$) | $a_{11} = 4.69(T-393)\cdot10^6 - 2.02\cdot10^8$, $a_{12} = 3.230\cdot10^8$, <br> ($a_{11} = -6.71\cdot10^8$ at 298 K) |
| $a_{ijk}$   (in m$^9$J/C$^6$) | $a_{111} = -5.52(T-393)\cdot10^7 + 2.76\cdot10^9$, $a_{112} = 4.47\cdot10^9$, $a_{123} = 4.91\cdot10^9$ <br> (at 298 K $a_{111} = 82.8\cdot10^8$, $a_{112} = 44.7\cdot10^8$, $a_{123} = 49.1\cdot10^8$) |
| $Q_{ij}$  (m$^4$/C$^2$) | $Q_{11}=0.11$, $Q_{12}= -0.043$, $Q_{44}=0.059$ |
| $s_{ij}$   (in $10^{-12}$ Pa$^{-1}$) | $s_{11}=8.3$, $s_{12}= -2.7$, $s_{44}=9.24$ |
| $g_{ij}$   (in $10^{-10}$m$^3$J/C$^2$) | $g_{11}=5.0$, $g_{12}= -0.2$, $g_{44}= 0.2$ |
| $F_{ij}$ (in $10^{-11}$m$^3$/C) <br> $f_{ij}$ (in V) | $F_{11} = 2.4$, $F_{12} = 0.5$, $F_{44} = 0.6$ (the first two values are recalculated from [a] values   $f_{11} = 5.1$, $f_{12} = 3.3$, $f_{44} = 0.065$ V. <br> The equality $F_{44} = F_{11} - F_{12}$ is valid in the isotropic case. |

[a] I. Ponomareva, A.K. Tagantsev, L. Bellaiche, Finite-temperature flexoelectricity in ferroelectric thin films from first principles, Phys. Rev. B **85** (2012) 104101.

Analytical calculations are performed and visualized in the Mathematica 12.2 software (https://www.wolfram.com/mathematica).

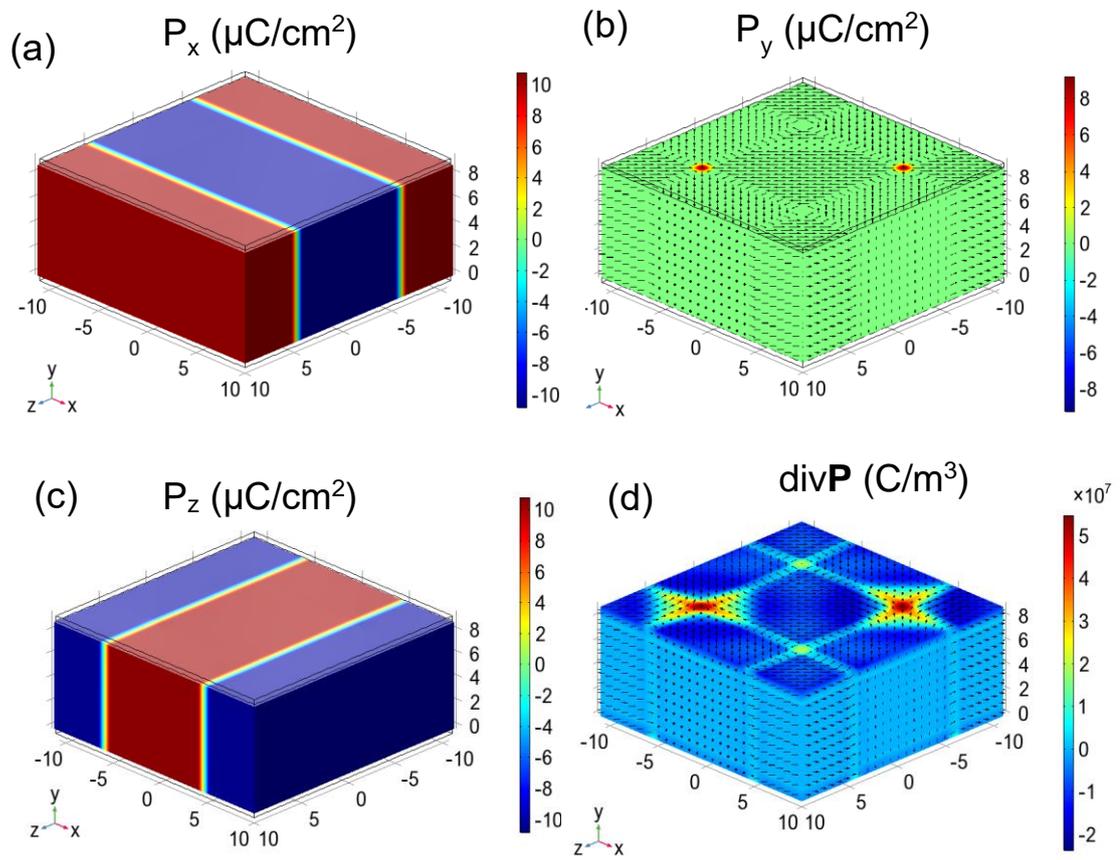

**Figure A1. (a)-(c)** Polarization components in the initial distribution, whose relaxation leads to the lateral grid of vortices and antivortices. Two red spots in the plot **(c)** are the seeding of antivortex cores. Arrows in the plot (c) show the polarization direction. The small random component superimposed on the regular structure is not shown. **(d)** Polarization divergence is the relaxed structure consisting of two vortex-antivortex pairs (final state).

**APPENDIX B. Supplementary Figures**

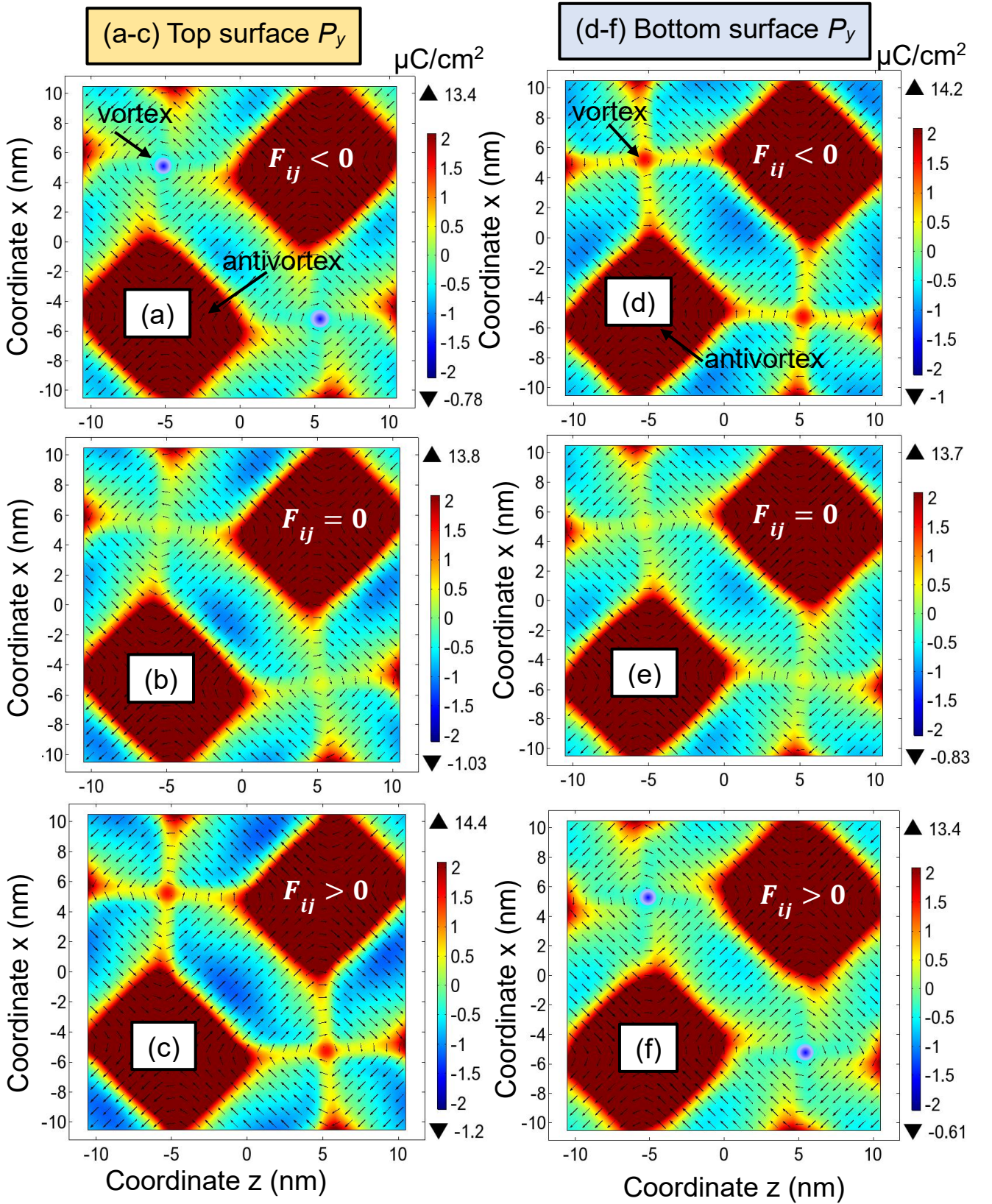

**Figure B1**. XZ-section of the out-of-plane polarization component $P_y$ at the film top (**a-c**) and bottom (**d-f**) surfaces calculated for negative (**a,d**), zero (**b,e**) and positive (**c,f**) flexoelectric coefficients $F_{ij}$. Arrows in the plots show the direction of polarization in the vortex-antivortex region. The thickness of BaTiO$_3$ film is 9 nm, mismatch strain $u_m$=0.1%, and the temperature is 300 K.

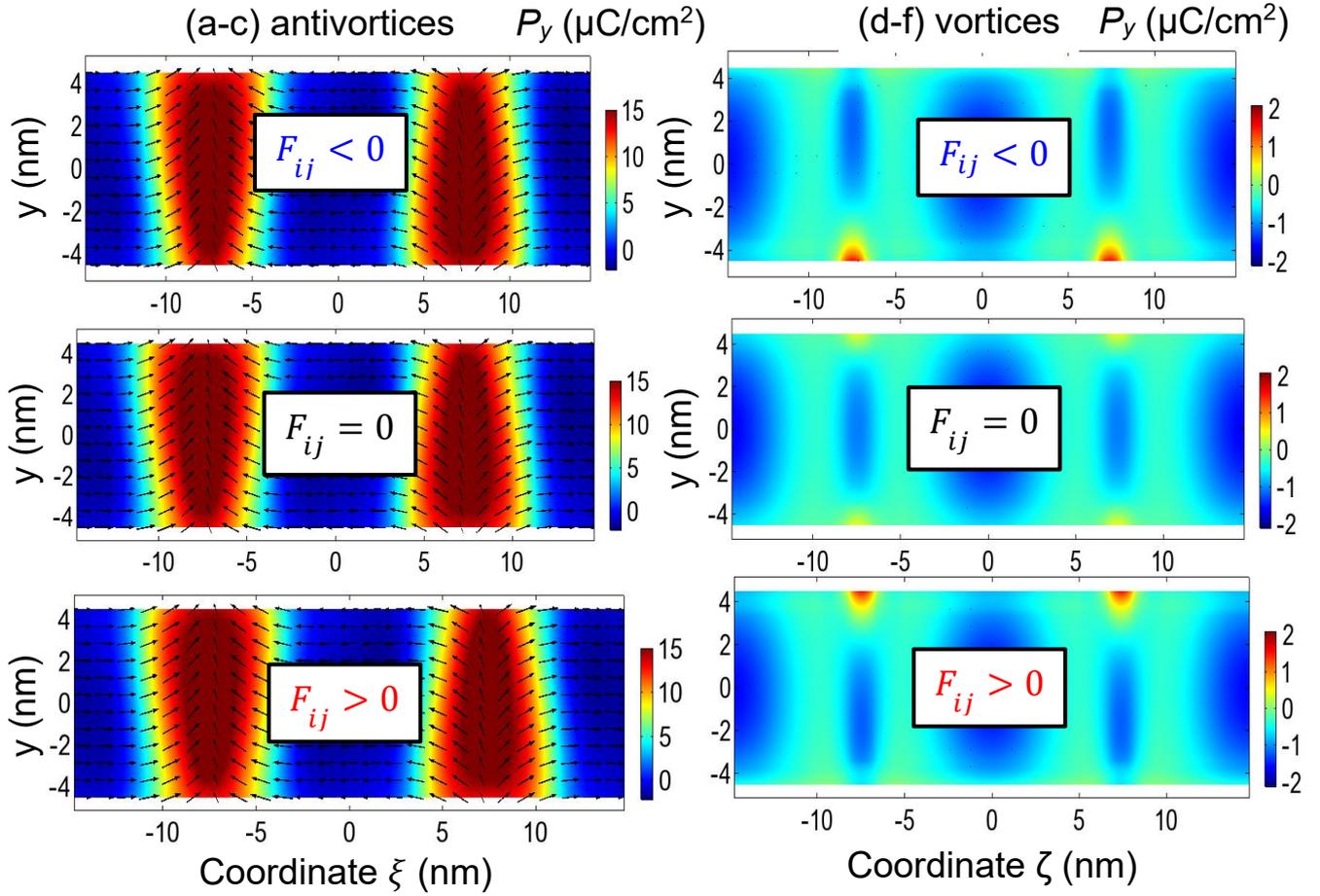

**Figure B2**. $\xi Y$-section **(a-c)** and $\zeta Y$-section **(d-f)** of the $P_y$ calculated for negative **(the top row)**, zero **(the middle row)** and positive **(bottom row)** flexoelectric coefficients $F_{ij}$. Coordinates $\xi = (x + z)/\sqrt{2}$ and $\zeta = (x - z)/\sqrt{2}$. Arrows in the plots (a)-(c) show the direction of polarization vector. The film thickness is 9 nm, mismatch strain $u_m$=0.1%, and the temperature is 300 K.

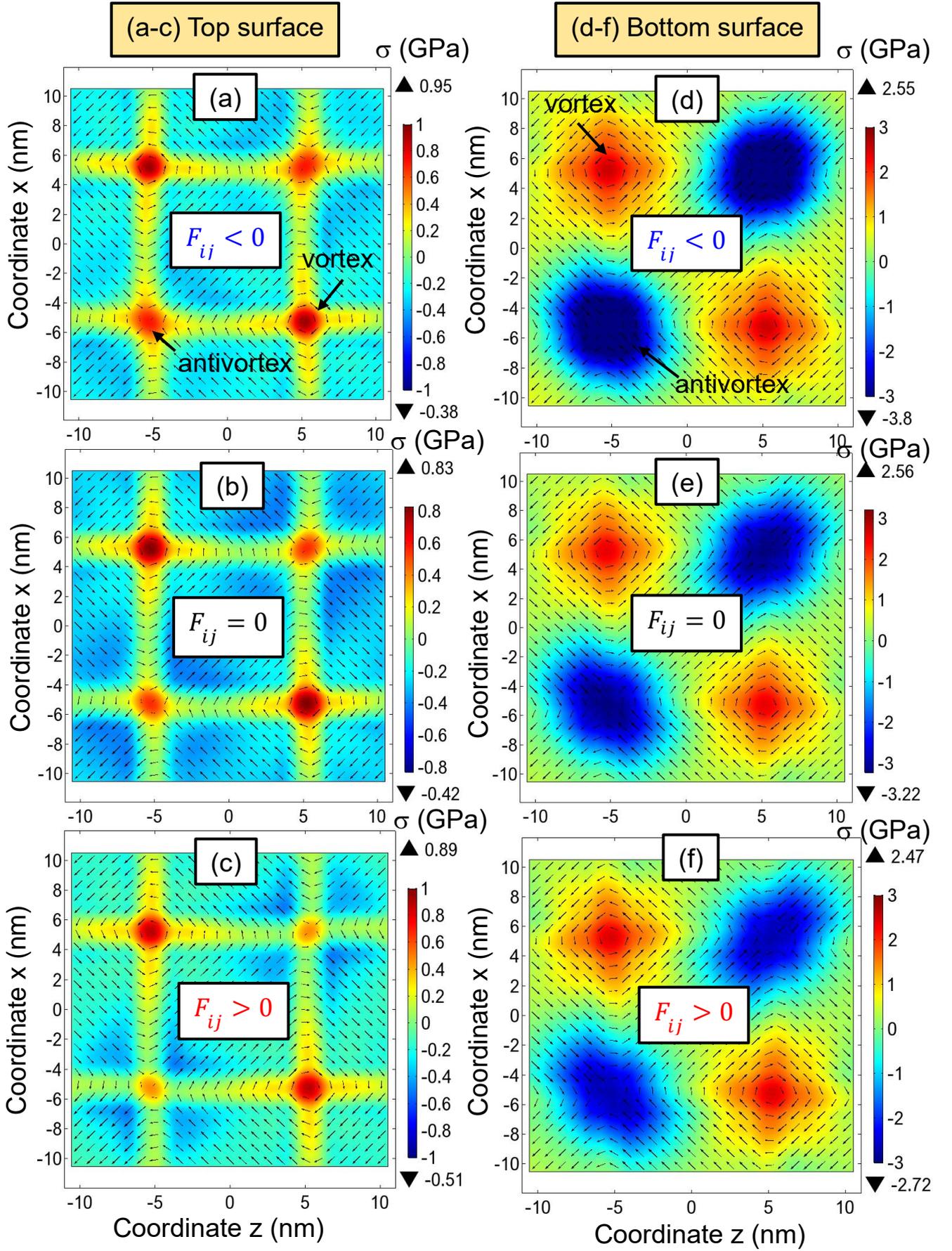

**Figure B3**. XZ-section of hydrostatic stress $\sigma$ at the film top **(a-c)** and bottom **(d-f)** surfaces calculated for negative **(a,d)**, zero **(b,e)** and positive **(c,f)** flexoelectric coefficients $F_{ij}$. Arrows in the plots show the direction of polarization vector. The film thickness is 9 nm, mismatch strain $u_m=0.1\%$, and the temperature is 300 K.